\documentstyle[12pt,aps,epsf,preprint,tighten]{revtex}

\def\bB{${\partial {\cal B}}$}
\def\iB{${\cal B}$}
\def\dE{$\langle |\Delta E|\rangle$}

\begin{document}
\draft
\title{
\begin{flushright}
{\small CNS-9730\\
November 1997\\
Phys. Rev. E {\bf 57}, 4095-4105(1998)}
\end{flushright}
Relevance of chaos in numerical solutions of quantum billiards
}
\author{Baowen Li$^{1,2}$, Marko Robnik$^{2}$, and Bambi Hu$^{1,3}$} 
\address{
$^{1}$ Department of Physics and Centre for Nonlinear Studies, Hong Kong 
Baptist University, Hong Kong, China \\
$^{2}$ Center for Applied Mathematics and Theoretical Physics,
University of Maribor, Maribor, Slovenia\\
$^{3}$ Department of  Physics, University of Houston, Houston TX 77204, USA
}
\date{\today} 
\maketitle

\begin{abstract} 

In this paper we have tested several general numerical methods in solving
the quantum billiards, such as the boundary integral method (BIM) and the
plane wave decomposition method (PWDM).  We performed extensive numerical
investigations of these two methods in a variety of quantum billiards:
integrable systens (circles, rectangles, and segments of circular
annulus), Kolmogorov-Armold-Moser (KAM) systems (Robnik billiards), and
fully chaotic systems (ergodic, such as Bunimovich stadium, Sinai billiard
and cardiod billiard). We have analyzed the scaling of the average
absolute value of the systematic error $\Delta E$ of the eigenenergy in
units of the mean level spacing with the density of discretization $b$
(which is number of numerical nodes on the boundary within one de Broglie
wavelength) and its relationship with the geometry and the classical
dynamics. In contradistinction to the BIM, we find that in the PWDM the
classical chaos is definitely relevant for the numerical accuracy at a
fixed density of discretization $b$.  We present evidence that it is not
only the ergodicity that matters, but also the Lyapunov exponents and
Kolmogorov entropy.  We believe that this phenomenon is one manifestation
of quantum chaos. 

\end{abstract} 

\pacs{PACS numbers: 05.45.+b, 02.70.Rw, 03.65.Ge}


\section{Introduction}
It is quite embarrassing to realize that in an attempt to numerically solve the
Helmholtz equation   
\begin{equation}
\nabla_{{\bf r}}^2 \psi({\bf r}) + k^2 \psi({\bf r}) = 0,
\label{eq:Helmholtz}
\end{equation}
satisfied by the scalar solution $\psi({\bf r})$ with eigenenergy $E=k^2$
inside  a connected plane domain ${\cal B}$ with the Dirichlet boundary
condition $\psi({\bf r})=0$ on the boundary ${\partial {\cal B}}$, one can face
enormous difficulties in cases of "\,problematic\," geometries such as 
various
nonconvex shapes. This is precisely the problem of solving and describing the
quantum billiard ${\cal B}$ as a Hamiltonian dynamical system, which is thus
just the two-dimensional  Schr\"odinger problem for a free point 
particle moving inside the
enclosure ${\partial {\cal B}}$, described by the wave function 
$\psi({\bf r})$ 
with the eigenenergy $E=k^2$. The corresponding classical problem is the
classical dynamics of a freely moving point particle obeying the law of specular
reflection upon hitting the boundary ${\partial{\cal B}}$.
Quantum billiards and their correspondence to their classical counterparts,
especially in the semiclassical level, are important model systems in studies of
quantum chaos\cite{Gutz90,GVZ91,CC95}.
There are several {\em general} methods for a numerical solution of
Eq.(\ref{eq:Helmholtz}) such as the boundary integral method (BIM) 
(see e.g.\cite{MK79,BW84,Boas94}),
and the plane wave decomposition method (PWDM), introduced and advocated by 
Heller\cite{Heller84}, whose analysis, especially in the light of the 
relevance of
classical chaos, is the subject of our present paper. Another quite general
method is the conformal mapping diagonalization technique introduced by 
Robnik\cite{Rob84} and further developed by Berry and 
Robnik\cite{BRob86}, Prosen
and Robnik\cite{PRob934}, and  Bohigas {\it et al.} \cite{Boh93}, 
which in principle should work 
for any shape, whereas, in
practice it is used for shapes for which the conformal mapping onto the unit
disk (or some other integrable geometries admitting a simple basis for
the representation) 
is sufficiently simple (possiblly also analytic). These methods can
face quite similar problems in cases of almost intractable geometries, but they
are to some extent complementary. For example, the conformal mapping
diagonalization technique can
provide a complete set of all eigenenergies up to some maximal value beyond
which the calculations cannot be performed due to the lack of computer storage
(RAM),  which means that we cannot reach very high-lying eigenstates.
(Our present record\cite{PRob934} is about 35~000 
for the size of
the banded matrix that we diagonalize in double precision, yielding at least 
12~000 good levels with accuracy of at least $1\%$ of the mean 
level spacing.) 

However, using the PWDM it is possible to go higher in energy by 
orders of
magnitude but then only a few selected states can be calculated with many
intermediate states in the spectral stretch missing. Therefore, the 
geometry of
some interesting and representative high-lying states can be analyzed,  but the
sample is typically not sufficiently complete (there are many states 
missing)  to 
perform statistical analysis. See, e.g., our recent papers on this 
topic\cite{LiRob94,LiRob95a}. The reasons for a failure of one of these 
methods  
can be quite different. For example, in the BIM the main difficulty stems 
from the
existence of "\,exterior chords\," in nonconvex geometries in its standard
formulation (see Sec. III), but the trouble might be overcome by an 
appropriate reformulation
of the method adapted to the correct semiclassical behaviour. We will 
discuss this in Section III where we also show that classical chaos is 
completely
irrelevant for the BIM. On the contrary, in the PWDM we find that the 
classical chaos
is relevant for numerical accuracy especially in the semiclassical limit of
the sufficiently small effective Planck constant $\hbar_{eff}$ 
reached at sufficiently
high eigenenergies. This demonstration and its qualitative explanation is the
main subject of our present paper. To give a specific example we should mention
isospectral billiards discovered and proved by Gordon {\em et 
al.}\cite{Gordon92},  which
have been investigated experimentally by Sridhar and Kudrolli\cite{SK94}  
and it is
also our experience\cite{LiRob95b} that the BIM fails in this
case (namely, due to strong nonconvexities), whereas the PWDM at $b=12$ 
yields the
accuracy of eigenenergies of within a few percent of the mean level
spacing, except for some very special eigenmodes for which surprisingly 
we find agreement
within double precision (16 digits) and which are characterized by the fact
that these eigenvalues agree with the analytic solutions for the triangles
within single precision (eight digits). So the fact that in this and similar 
cases
the experimental precision (for some levels) exceeds the best possible 
numerical precision even when using the best available methods is embarrassing 
for a theoretician, but also motivation for further work.      

The paper is organized as follows. In Sec. II, we focus our 
attention on the PWDM. In Sec. III, 
we first point out some serious 
flaws in the derivation of the BIM in the literature and show how the final 
formula (which nevertheless was correct) should be derived in a 
{\em regularized} way and then discuss the 
numerical results of the BIM and the relevance with classical dynamicss and 
geometry, etc. In Sec. IV we give a discussion and conclusions.

\section{The plane wave decomposition method of Heller}

\subsection{The numerical procedure of the PWDM}

In this section we present our general exposition of PWDM 
following\cite{LiRob94}. To solve the Schr\"odinger equation  
(\ref{eq:Helmholtz})  
for $\psi({\bf r})$ with the Dirichlet boundary condition $\psi({\bf r}) 
=0$ on 
\bB $\;$ we use the ansatz of the superposition  of plane waves 
(originally due to Heller\cite{Heller84})
\begin{equation}
\psi({\bf r}) = \sum_{j=1}^{N} a_{j} \cos (k_{xj}x + k_{yj}y + \phi_j),
\label{eq:pwd}
\end{equation}
where $k_{xj} = k\cos\theta_j,\quad k_{yj} = k\sin\theta_j,\quad k^2 = E$,
and we use the notation ${\bf r} = (x,y)$. $N$ is the number of plane waves 
and $\phi_j$ are {\em random phases}, drawn from the 
interval $[0,2\pi)$, assuming a
uniform distribution, and $\theta_j=2j\pi/N$ determining the direction angles of
the wave vectors chosen equidistantly. The ansatz (\ref{eq:pwd}) solves
the Schr\"odinger equation (\ref{eq:Helmholtz}) in the interior of the billiard
region \iB, so that we have only to satisfy the Dirichlet boundary condition. 
Taking the random phases, as we discovered, is equivalent
to spreading the origins of plane waves all over the billiard region,
and at the same time this results in reducing the CPU time by
almost a factor of 10.
For a given $k$ we set the wave function equal to zero at 
a finite number $M$ of boundary points (primary nodes) and 
equal to 1 at an arbitrarily chosen interior point. Of course, $M\ge N$.
This gives an inhomogeneous set of equations that can be solved 
by matrix inversion.  Usually the matrix is very singular and  thus the 
{\em singular value decomposition} method has been 
invoked\cite{Heller84,Press92}.  
After obtaining the coefficients $a_{j}$ we 
calculate the wave functions at other boundary points (secondary nodes). 
We always have three secondary nodes between a pair of primary nodes. The
experience shows that a further increase of the number of secondary nodes 
does not enhance the accuracy.
The sum of the squares of the wave function at all the secondary nodes
(Heller called this sum "\,tension\,") would  be ideally zero if $k^2$ is an
eigenvalue and if Eq.(\ref{eq:pwd}) is the corresponding exact solution of
Eq.(\ref{eq:Helmholtz}). 
In practice it is a positive number. Therefore, the eigenvalue
problem now is to find the minimum of the tension. In our numerical
procedure we have looked for the zeros of
the first derivative of the tension; namely, the derivative is available 
analytically or explicitly from Eq.(\ref{eq:pwd}) once the amplitudes $a_j$
have been found.  In this paper we make the choice $M=N$ since it proved 
to be sufficient for calculating the lowest 100 states whose accuracy we
analyze. (For the high-lying states studied in our paper\cite{LiRob94},  
we have used $M = 5N/3$.) 
It must be pointed out that the wave functions obtained in 
this way are not (yet) normalized due to the arbitrary choice of the 
interior point where the value of the wave function has been arbitrarily
set equal to unity. We therefore explicitly normalize these wave functions 
before embarking on the analysis of their properties. 

The accuracy of this method of course depends on the number of plane waves
($N$) and on the  number of the primary nodes ($M$) and we have a considerable
freedom in choosing $N$ and $M\ge N$.
In order to reach a sufficient accuracy the experience shows that
we should take at least $N=3{\cal L}/\lambda_{de~Broglie}$ and $M=N$, 
where 
${\cal L}$ is the perimeter of the billiard and $\lambda_{de~Broglie}$
is the de Broglie wavelength equal to $2\pi/k$. With this choice in the 
present context and for the lowest 100 states we reach the double
precision accuracy (16 digits) for all levels of integrable
systems such as the rectangular billiard (where the eigenenergies can be
given trivially analytically) and the circular billiard, but also
for the Robnik billiard  ${\cal B}_{\lambda}$ for small $\lambda \le 0.1$. 
Introducing the density of discretization $b$ defined as the number of
numerical nodes per one de Broglie wavelength on the boundary we thus write the
number of plane waves $N = b {\cal L}/\lambda_{de~Broglie} = b2\pi{\cal 
L}/k$.

The main problem of investigation in this paper is to study the dependence of
the systematical numerical error $\Delta E$ (i.e., the error due to the 
finite discretization) on the density of discretization
$b$ and the dependence of $\Delta E$ on the geometry (billiard shape
parameter) at fixed $b$. In order to perform a systematic analysis the errors
should be measured in some natural units and in our case this is of course just
the mean level spacing, which, according to the leading term of the Weyl 
formula, is
equal to $4\pi/{\cal A}$, where ${\cal A}$ is the area of the billiard 
\iB. From
now on we shall always assume that $\Delta E$ of a particular energy level is
in fact measured in such natural units. 
Of course one immediately realizes that the error $\Delta E$ fluctuates
wildly from state to state (see Figs. 5 and  7) 
so that generally nothing can be predicted about it individually.
Therefore, the approach must be a statistical one and so we typically
take an average of the errors $\Delta E$ over a suitable ensemble of states.
Specifically, in all cases of this paper we have taken the average of the
absolute values of $\Delta E$ over the lowest 100 states (of a given symmetry
class) and denoted it by \dE. 
It is important and should be mentioned that we have also
checked the stationarity of such an average value over consecutive
spectral stretches of 100 states each, so that our procedure does make sense. 
In addition, we have also investigated the standard deviation  
$\sigma_{|\Delta E|} =\langle (|\Delta E| - 
\langle |\Delta E|\rangle)^2\rangle^{1/2}$, which always 
has the same order of magnitude as the average value.

It turns out that the accuracy of energy levels depends nontrivially on $b$,
unlike in the BIM, where we find always a power law (see Sec. III),
namely, it typically shows broken power law. By this we mean that \dE 
~obeys a power law
\begin{equation}
\langle|\Delta E|\rangle = A b^{-\alpha},
\label{eq:power-law}
\end{equation}
with very large $\alpha$ for sufficiently small $b$,
$b\le b_c$, whereas for larger $b\ge b_c$ it obeys a rather flat power law with
very small positive $\alpha$ (close to zero). Therefore, in contradistinction
to the BIM, it
is difficult to explore the general dependence of \dE~ on $b$ if there 
is any
such universality at all. However, in order to investigate the dependence of
the accuracy on geometry and the implied dynamical properties of billiards,
we have
decided to fix the value of $b$ and  have chosen $b=12$, and then we look at 
the dependence of \dE~ on the shape parameters of three one-parameter 
billiards, namely, the Robnik billiard, the Bunimovich stadium, and the 
Sinai billiard.

Finally, we would like to discuss how to estimate the error.  As 
usual, to speak of an error we need to have a standard value. The 
question is 
how to get the standard values in different quantum billiards.  As for the 
integrable billiard, such as a
rectangle, we know analytically the exact values.
For a circular billiard, they are the zeros of the Bessel 
function, which can be calculated very precisely. However, for other 
billiards, in particular the chaotic billiards, there are no
true accurate values available (otherwise we do not need the numerical 
methods anymore).
In fact, both the PWDM and the BIM can be self-tested for their accuracy. On 
the one hand, in both 
cases,  the numerical value at very large $b$ can be regarded as the 
"\,true\," 
value. On the other hand, in the PWDM one may change the position of the  
interior point and compare the two lists of eigenenergies obtained.
Moreover, since we have also some other special methods invented for the
billiard of a specific geometry, such as the diagonalization method for 
the Robnik billiard and the scattering approach for the Sinai billiard, in 
these techniques the accuracy is well controlled and we may obtain more 
accurate results than the BIM and PWDM; thus we can use the eigenenergy 
list from them as the standard value.

In our studies in this paper, the "\,standard value\," of the Robnik 
billiard are provided by Prosen\cite{Prosen956} by using the 
diagonalization 
method with a very large dimension of the matrix and thus the lowest 1000
eigenvalues are guaranted with an accuracy of at least $10^{-12}$ in 
units of the mean level spacing for the large shape parameter $\lambda$. The 
eigenvalues of the Sinai billiard were 
provided by Schanz and Smilansky\cite{Schanz95} by using their scattering 
method. The accuracy is about $10^{-7}$ of the mean level 
spacing, which is already high enough for our purpose. For the 
billiards whose eignevalues are not available from other methods, we always 
take the 
eigenvalues at very large density of discretization $b$ (say 30) as the 
true value.

\subsection{Relevance of chaos with the numerical accuracy}

As is well known,  in the classically integrable 
quantum Hamiltonian 
systems in the semiclassical limit (of sufficiently small $\hbar$)
the eigenfunction can locally be described by a {\em finite}
superposition of plane waves with the same wave number; in the case of plane
billiards it is  $k=\sqrt{E}$. If the quantum system has ergodic classical
dynamics then in the semiclassical limit locally the wave function can be
represented as a superposition of {\em infinitely} many plane waves with the
same $k$ and with the wave vectors being isotropically distributed on the
circle of radius $k$\cite{Berry77}. Moreover, the ergodicity suggests to 
assume random phases
for the ensemble of plane waves, which implies that to the lowest 
approximation
the wave function is a Gaussian random function. While this is a good 
starting
approximation, originally due to Berry\cite{Berry77} and recently 
verified by Aurich
and Steiner\cite{AuSt93} and also by Li and Robnik\cite{LiRob94}, the 
phases are actually
not random but correlated in a subtle way dictated by the classical dynamics,
especially along the short and the least unstable periodic orbits, which is the
origin of the scar 
phenomenon\cite{Heller84,Bog88,Berry89,Li97,LiHu97}. 
Thus we can qualitatively very well
understand that the PWDM should work well or even brilliantly in cases of
classically integrable billiards whereas in the ergodic systems we expect a
severe degradation of the accuracy (at fixed $b$) simply because the finite
number of plane waves cannot capture the correct (infinite) superposition of
plane waves everywhere in the interior of the billiard. If the system is 
a generic system of a mixed type with regular and irregular regions coexisting
in the classical phase space, a scenario described by the 
Kolmogorov-Arnold-Moser (KAM) theory, then the
degradation of accuracy (at fixed $b$) with increasing fractional measure of 
the chaotic component (denoted by $\rho_2$) is certainly expected. However,
$\rho_2$ is not the only parameter that controls the accuracy (at fixed $b$)
since, as we shall see, the dynamical properties such as the diffusion time, 
Lyapunov
exponent, and Kolmogorov entropy also play a role. It is the aim of the 
present paper to numerically explore this type of behavior in three different 
billiard systems.

The first billiard system is defined as the quadratic (complex) conformal map
$w = z + \lambda z^2$ from the unit disk $|z|\le 1$ from the $z$ plane onto the 
$w = (x,y)$ complex plane. The system has been introduced by 
Robnik\cite{Rob83} and further studied by Hayli {\em et al.}\cite{Hayli87}, 
Frisk\cite{Frisk90}, and Bruus and Stone\cite{BrSt934}
for various parameter values $\lambda$.
Since the billiard (usually called the Robnik billiard by the people in the 
community) has an analytic boundary it
goes continuously from the integrable case (circle, $\lambda=0$) through a 
KAM-like
regime of small $\lambda\le 1/4$ with mixed classical dynamics and becomes
nonconvex at $\lambda=1/4$ (the bounce map becomes discontinuous), where the
Lazutkin caustics (invariant tori) are destroyed giving way to 
ergodicity. It was
shown by Robnik\cite{Rob83} that the classical dynamics at these values of 
$\lambda$ is
predominantly chaotic (almost ergodic), although Hayli {\em et 
al.}\cite{Hayli87} have
shown that there are still some stable periodic orbits surrounded by very tiny
stability islands up to $\lambda=0.2791$. At larger $\lambda$ we have
reason and numerical evidence\cite{LiRob96a} to expect 
that the 
dynamics can be ergodic. Recently, it has been proven rigorously by 
Markarian\cite{Markar93} that for $\lambda =1/2$ (a cardioid billiard) the 
system is indeed
ergodic, mixing, and K. This was a further motivation to study the cardioid
billiard classically, semiclassically, and quantanlly by several groups,
e.g.,
B\"acker {\em et al.}\cite{Backer945}, and Bruus and Whelan\cite{BrWh96}. 
The billiard shape for
$\lambda =0.4$ is  shown (the upper half) in Fig. 1(a). Since all states are
either even or odd we can take into account these symmetry properties
explicitly. In fact, we want to specialize to the odd eigenstates only.
Therefore, in order to {\it a priori} satisfy the Dirichlet boundary 
condition on the
abscissa of Fig. 1a we specialize the general ansatz (\ref{eq:pwd}) to
the form
\begin{equation}
\psi({\bf r}) = \sum_{j=1}^{N} a_j \cos (k_{xj}x + \phi_j) \sin(k_{yj}y),
\label{eq:robnik}
\end{equation}
where all the quantities are precisely as in Eq.(\ref{eq:pwd}) except that 
the $N$ discretization (primary) nodes are equidistantly located only
along the half of the full billiard boundary, so that $b$ is exactly the same
as in using the ansatz (\ref{eq:pwd}) for the full billiard. 

In Fig. 2 we show the results for this billiard, namely we plot \dE~  (in 
logarithmic units) versus
$\lambda$ at fixed $b=12$. Close to integrability ($\lambda\le0.1$) we reach
the accuracy within 14--15 digits, which is almost the double precision on 
our machine (16 digits), in which all our calculations have been performed.
As the value of $\lambda$ increases we observe a dramatic deterioration of the
accuracy where \dE~ increases by many orders of magnitude, namely, by 
almost 
13 decades, leveling off at \dE~ approximately equal to $10^{-2}$, which 
means
that we have now the accuracy of only a few percent of the mean level spacing.
This dramatic but quite smooth increase of \dE~ is certainly related to the
emergence of classical chaos with increasing $\lambda$, but definitely is
{\em not} controlled merely by $\rho_2$ because $\rho_2$ reaches the value of
1 (almost ergodicity) already at $\lambda =1/4$\cite{PRob934,Rob83},  
whereas \dE~ still varies considerably in the region $\lambda \ge 1/4$.
Thus it is obvious that in the semiclassical picture also other classical
dynamical properties (measures of the "\,hardness\," of chaos) play an 
important 
role. Although we do not have a quantitative theory yet one should observe that 
according to Robnik\cite{Rob83} the Lyapunov exponent and Kolmogorov 
entropy ($h$)
vary also quite smoothly with $\lambda$, suggesting a speculation that there 
might be a relation between \dE~ and $h$.

Another demonstration of the effectivity of the PWDM and its accuracy is 
displayed
in Table I where we show the numerical value of the scalar product of two
consecutive normalized eigenstates, namely, the ground state and the first
excited state, denoted by $O_{12}$, which ideally should be zero. We see 
here too that the accuracy decreases (by orders of magnitude)
sharply but smoothly with increasing shape parameter $\lambda$.

It is then interesting to similarly analyze an ergodic system such as 
the stadium
of Bunimovich shown in Fig. 1(b), where the shape parameter is $a/R$ and we
have looked at the results for $0\le a/R \le 10$. In fact, for our 
purposes we
have chosen and fixed $R=1$ in all cases. Since this billiard is known
to be rigorously ergodic (and mixing and $K$) for any $a > 0$ in this case
$\rho_2$ is exactly 1 and constant. 
We have calculated the lowest 100 energy levels of the odd-odd symmetry class.
Therefore, in this case the general ansatz (\ref{eq:pwd}) can be
specialized as 
\begin{equation}
\psi({\bf r}) = \sum_{j=1}^{N} a_j \sin(k_{xj}x)\sin(k_{yj}y).
\label{eq:stadium}
\end{equation}
Here again the discretization (primary) nodes are only on the outer boundary of
the stadium with discretization density $b=12$.  
From our plot in Fig. 3 we see that in the
integrable case of the circle ($a =0$) we again reach the accuracy within at
least 14 digits, but this brilliant accuracy at fixed $b=12$ deteriorates
almost discontinuously upon increasing $a$ and then \dE~ still increases by
about two orders of magnitude when $a$ goes from 0.1 to 10. It appears to us
that classical chaos is definitely relevant for the accuracy of the method
which might and should be explained by an appropriate theory in the
semiclassical level. As an observation we should mention that the Kolmogorov
entropy increases sharply with $a/R$ when $a/R$ goes from 0 to about 1, where
it reaches the maximum, and then decreases slowly\cite{BeSt78},
whereas our \dE~ increases monotonically. Thus if there is a relationship 
between \dE~ and Kolmogorov entropy it certainly is not a simple one.  

We have tested also another system with hard chaos, namely, the Sinai 
billiard
sketched in Fig. 1(c) (desymmetrized). The system is known to be ergodic,
mixing, and $K$. In calculating the 100 lowest energy levels of the 
desymmetrized Sinai billiard we used the same specialized ansatz as in
Eq. (\ref{eq:robnik}), thereby taking into account explicitly the Dirichlet
boundary condition on the abscissa $y=0$. In this case $b$ is the density of
discretization of the equidistant nodes along the rest of the perimeter. 
Similarly as in the case of the stadium we easily reach the double
precision of 16 digits in the limiting integrable case of zero radius $R=0$,
but this accuracy is almost instantly lost by increasing $R$, as seen in Fig.
4. The value of \dE~ levels off at about $10^{-4}--10^{-2}$ for all $R$
between $0.025$ and $0.45$.   

As a final technical point we comment on the stationarity of \dE~ as a 
function
of energy, which has been confirmed for the Robnik billiard at 
$\lambda=0.27$,
where the average value over consecutive spectral stretches over 100 states has
been found to be quite stable. 
Specifically, to illustrate this finding we plot in Fig. 5 the absolute 
values of the errors of the lowest 400 consecutive eigenstates where one 
can see that 
the average value over 100 consecutive states is quite 
stable 
indeed. This is shown in Table II for four intevals of 100 states each.
We should emphasize again that the fluctuation is very big. 
We have calculated the standard deviation for 
these 400 levels; it is $\sigma_{|\Delta E|}= 7.40\times 10^{-7}$, which 
has the same order of magnitude as the average value (see Table II).
It is our numerical experience that for all cases that the standard 
deviations are always about the same order of average values.

\section{Boundary integral method}

After discussing the PWDM, we now turn to another very important 
numerical method: the boundary integral method. This method is 
widely used 
not only in studying quantum chaos, but also in 
engineering\cite{Baner94}.
In this section we would 
give an extensive numerical investigation of its accuracy in relation to 
classical dynamics and geometrical properties.
However, before we go into a detailed numerical and technical analysis, we 
would like to point out two serious flaws in the derivation of the BIM in 
the literature and show how the final formula, which nevertheless is 
correct, should be derived in a sound way.

\subsection{A regularized derivation of the BIM}

In order to clearly expose the difficulties and the errors in the derivation
of the BIM offered in the literature, e.g. in Ref\cite{BW84}, we present 
our {\em regularized} derivation, by which we mean that we construct and 
use a Green
function that automatically (by construction) satisfies the Dirichlet 
boundary
condition (vanishes locally on the boundary $\partial {\cal B}$ of the 
billiard domain
${\cal B}$), which is achieved by employing the method of images (see, e.g.,
the article of Balian and Bloch\cite{BB74} and the references therein). 
This will enable us to avoid committing two
errors, which, however, luckily compensated for each other. {\em First}, 
in taking
the normal derivatives on the  two sides of Eq. (6) in Ref\cite{BW84}, on 
the right hand side we must use the
value $\psi({\bf r})$ which is the interior solution inside ${\cal B}$,
rather than $\frac{1}{2}\psi({\bf r})$, which is the value exactly on the
boundary, simply because in taking the derivatives we must evaluate the
function at two infinitesimally separated points normal to the boundary. {\em
Second}, this error of taking the unjustified factor $1/2$ is then
exactly compensated for by another error in arriving at Eq. (8) in 
Ref. \cite{BW84},
namely, by interchanging the integration along the boundary $\partial 
{\cal B}$ 
and the normal differentiation, because due to singularities on the boundary 
these two operations do not commute.

Now, we offer our regularized derivation. We are searching for the solution
$\psi({\bf r})$ with eigenenergy $E = k^2$ obeying the Helmholtz equation 
(\ref{eq:Helmholtz}), with the Dirichlet boundary condition $\psi({\bf 
r})=0$ on the boundary
 ${\bf r} \in \partial {\cal B}$.
We will transform this Schr\"odinger equation for our quantum billiard ${\cal
B}$ into an integral equation by means of the {\em regularized} Green function 
$G({\bf r}, {\bf r}^{\prime})$, which solves the defining equation
\begin{equation}
\nabla_{{\bf r}}^2 G({\bf r},{\bf r'})  + k^2 G({\bf r},{\bf r'}) 
= \delta({\bf r}-{\bf r'}) - \delta({\bf r}-{\bf r'}_{\it R}),
\label{eq:Green-eq}
\end{equation}
where ${\bf r}$ and  ${\bf r'}$ are in $ {\cal B}\cup {\partial {\cal B}}$
and ${\bf r'}_{\it R}$ is the mirror image of ${\bf r'}$ with respect to the
tangent at the closest-lying point on the boundary (and thus if ${\bf r'}$ 
is sufficiently close to the boundary then ${\bf r'}_{\it R}$ is outside the
billiard ${\cal B}$) (see Fig. 6).
The solution can easily be found in terms of the free propagator (the 
free-particle Green function on the full Euclidean plane) 
\begin{equation}
G_0({\bf r}, {\bf r'}) = - \frac{1}{4} i H_{0}^{(1)}(k|{\bf r}-{\bf r'}|),
\label{eq:Gfree}
\end{equation}
where $H_{0}^{(1)}$ is the zeroth order Hankel function of the first 
kind\cite{AS72}, namely, 
\begin{equation}
G({\bf r},{\bf r'}) = G_{0}({\bf r},{\bf r'}) -G_{0}({\bf r},{\bf r'}_{\it R}),
\label{eq:Green}
\end{equation}
such that now $G({\bf r},{\bf r'})$ is zero by construction for any
${\bf r'}$ on the boundary $\partial {\cal B}$, in contradistinction to the
Green function defined and used in Eq. (5) in \cite{BW84}. 
Multiplication of Eq.(\ref{eq:Green-eq}) by $\psi({\bf r})$
and the Helmholtz equation (\ref{eq:Helmholtz}) by $G({\bf r}, {\bf 
r'})$,
subtraction, integration over the area inside ${\cal B}$, and using Green's
theorem yields
\begin{equation}
\oint ds\left[\psi({\bf r}){\bf n}\cdot\nabla_{{\bf r}} G({\bf r},{\bf r'})
-G({\bf r},{\bf r'}){\bf n}\cdot\nabla_{{\bf r}} \psi({\bf r}) \right]
= \psi({\bf r'}),
\label{eq:BIM1}
\end{equation}
where $s$ is the arclength on the boundary $\partial {\cal B}$ oriented
counterclockwise, ${\bf n}$ is the unit normal vector to $\partial {\cal 
B}$ at ${\bf r}$
oriented outward, and this equation is now valid for {\em all} ${\bf r'}$ inside
and on the boundary of ${\cal B}$. Since in this equation everything is 
regular, we
can take the normal partial derivatives on both sides. Following the usual
notation in \cite{BW84} we define the normal derivative of $\psi$ at the 
point $s$ as \begin{equation}
u(s) = {\bf n}\cdot\nabla_{{\bf r}}\psi({\bf r}(s))
\label{eq:nder}
\end{equation}
and thus using the boundary condition $\psi({\bf r}) =0$ we arrive at
\begin{equation}
u(s) = -2\oint ds'u(s'){\bf n}\cdot\nabla_{{\bf r}}G_{0}({\bf r},{\bf r'}).
\label{eq:BIM2}
\end{equation}
In this way we have correctly derived the main integral equation of the
boundary integral method, which is correctly given as Eq. (8) in \cite{BW84} 
(where
the two errors exactly compensate for each other), so that all the further 
steps in working out
the geometry of Eq. (\ref{eq:BIM2}) and the numerical discretization are
exactly the same as in \cite{BW84}. As shown in Fig. 6, we define the length 
of the
chord between two points on the boundary ${\bf r}(s)$ and ${\bf r}(s')$ as
\begin{equation}
\rho(s,s') = |{\bf r}(s) -{\bf r}(s')|
\label{eq:chord}
\end{equation}
and the angle $\theta(s',s)$ is the angle between the chord and the tangent to
$\partial {\cal B}$ at $s'$. The chord is an oriented separation vector 
pointing from ${\bf r}(s)$ to ${\bf r'} = {\bf r}(s')$.
Of course $\theta(s,s') \neq \theta (s',s)$.
Thus, following the notation of \cite{BW84} we can write
\begin{equation}
{\bf n'}\cdot\frac{\partial G_0({\bf r},{\bf r'})}{\partial {\bf r'}} 
= \sin\theta(s',s)\frac{\partial G_0}{\partial \rho}
\label{eq:normgreen}
\end{equation}
and we obtain finally  
\begin{equation}
u(s) = -\frac{1}{2} ik\oint ds' u(s')\sin\theta(s,s')H_1^{(1)}\{k\rho(s,s')\}.
\label{eq:bim3}
\end{equation}
In numerically solving this integral equation we have used precisely the same
primitive discretization procedure as \cite{BW84}, which turned out to be 
better than
some other more sophisticated versions. So we simply divide the perimeter 
${\cal L}$ into $N$ equally long segments and thus define
\begin{equation}
s_m = m{\cal L}/N,\quad \rho(s_l,s_m) =\rho_{lm},\quad
\theta(s_l,s_m)=
\theta_{lm},\quad 
1\le l,m \le N.
\label{eq:notation}
\end{equation}
Therefore, numerically, we are searching for the zero of the determinant
$\Delta_N(E)=$det$(M_{lm})$, where $M_{lm}$ are the matrix elements 
of the $N\times N$ matrix
\begin{equation}
M_{lm} = \delta_{lm} + \frac{ik{\cal L}}{2N} \sin\theta_{lm}
H_1^{(1)}(k\rho_{lm}),
\label{eq:Mmatrix}
\end{equation}
where $E=k^2$. Due to the asymmetry $\theta_{lm} \neq \theta_{ml}$ this matrix
is a general complex non-Hermitian matrix. For the diagonal elements $l=m$,
where $\theta_{lm}$ is either zero or $\pi$, the proper limit of the second
term on the right hand side in Eq. (\ref{eq:Mmatrix})  must be taken and 
then it is equal to 
$\kappa(s){\cal L}/2\pi N$, where $\kappa(s)$ is the curvature of the 
boundary at $s$.

One important aspect of this formalism is the semiclassical limiting form 
that has been extensively studied by Boasman\cite{Boas94}. At this point we 
want to make 
the following comment. In the cases of nonconvex geometries we will have {\em
exterior} chords connecting two points on the boundary such that they lie
entirely, or at least partially,  outside ${\cal B}$. While formally the method
and the procedure in such cases is perfectly correct,
in reality it might be problematic, which
can be seen by considering the semiclassical limit. The
formal leading order in the asymptotic expansion of the Hankel function
$H_1^{(1)}$ (Debye approximation) does not match the actually correct
semiclassical leading approximation that would be spanned by the shortest
classical orbit connecting the two points via at least one or many collision
points in between. Therefore, we understand and expect that the method must 
meet some difficulties in cases of nonconvex geometry. This has 
been partially  confirmed in our present work as we will show in Sec. 
IIIB,  while the analytical work to
reformulate the method including the multiple collision expansions is in
progress\cite{LiRob95c} and is expected to deal 
satisfactorily with nonconvex geometries.

In the Sec. IIIB we shall analyze the numerical accuracy of the  BIM as a
function of the density of discretization 
\begin{equation}
b = \frac{2\pi N}{k{\cal L}},
\label{eq:defb}
\end{equation}
in a variety of quantum billiards with integrable, KAM-type, or ergodic
classical dynamics, including such with nonconvex geometry. The main result is
that there is always a power law so that the error of eigenenergy in units of
the mean level spacing, after taking the average of the absolute value 
over a 
suitable ensemble of eigenstates,  obeys $\langle|\Delta E|\rangle = A 
b^{-\alpha}$, 
but the exponent $\alpha$ (and the prefactor $A$) is nonuniversal.

\subsection{Numerical results} 

The numerical procedure we have used to solve the variety of quantum billiards
is exactly as described above and therefore it is 
precisely the same as in \cite{BW84}. Our main task now is to analyze in 
detail the behavior
of the BIM as a function of the density of discretization $b$, especially in
relation to the geometrical properties of ${\cal B}$ (nonconvexities) and in
relation to classical dynamics, whose chaotic behavior is expected to imply
interesting methodologic and algorithmic manifestation of quantum chaos.

Again, like in the PWDM, we have to measure the numerical error of 
the eigenenergies in units of the mean level spacing and perform some
kind of averaging over a suitable ensemble of  consecutive states. However, 
this
will make sense only if such a local average of the error is stationary
(constant) over a suitable energy interval. This condition has been confirmed
to be satisfied  in almost all cases that we checked. In the case of the 
circle
billiard and in the case of the cardioid billiard this stationarity of the 
locally 
averaged error is shown in Figs. 7(a-b), respectively, where we plot the data
for about $95\%$ of the lowest 1000 odd levels.
The average value \dE~ for these two cases are $3.89\times 10^{-5}$ (
circle) and $1.99 \times 10^{-3}$ (cardiod billiard), while the standard 
deviations $\sigma_{|\Delta E|}$ are $3.67\times 10^{-5}$ (circle)  and 
$2.91\times 
10^{-3}$ (cardiod billiard), respectively. Again, like in the case of 
the PWDM, 
the standard deviation in the BIM is of the same order as the average value.

We have established that we have always taken the average of the 
absolute value of the error (in units of the mean level spacing) 
over a suitable ensemble of eigenstates, for which we have chosen
the lowest 100 eigenstates in all cases. Actually, strictly 
speaking,
we have taken about 90--95 levels from the ensemble of the lowest 100 odd
levels: since in using the BIM we always miss some levels, the quota of 
missing levels typically is a few percent, depending on the step size.  
In our case the step size is 1/20 of the mean level spacing and the fraction of
missing levels varied from about 10 percent in integrable billiard with Poisson
statistics to about 5 $\%$ in ergodic cases with GOE statistics.

In Figs. 8(a-f) we show the error $\langle|\Delta E|\rangle$
versus $b$ for six billiard shapes. In Fig. 8(a) we have the full circle 
billiard.
In Fig. 8(b) we have 
$270^0$ ($=3\pi/2$) segment of the annulus
billiard with inner radius $R_1 = 0.45$ and outer radius $R_2 =0.5$. 
These data were based on the very careful work of Thomas 
Hesse\cite{Hess95}, who
has kindly communicated to us his unpublished results and the analysis, which
we have independently  checked  and confirmed. In 
Figs. 8(c) and (d) we have the Robnik billiard with shape parameters 
$\lambda =0.15$ and $\lambda =1/2$, respectively.
In Fig. 8(e) we show the results for the 1/4 Bunimovich stadium 
with the size $2\times 2$ for the central square. In Fig. 8(f) we have the 
1/4 Sinai
billiard with dimensions $2\times 2$ for the square and circular radius 
$R=1/2$. The best-fitting power-law curve is described by 
Eq.\ref{eq:power-law}
and is seen to provide a very significant fit in all six cases.

As for the Robnik billiard, one 
should be reminded that at $\lambda =0$ we have integrable classical
dynamics in the circle billiard and at $\lambda =0.15$ we have 
KAM-type dynamics with islands of stability\cite{LiRob95b}; it 
should
be emphasized that at $\lambda=1/4$ we have a zero curvature point at 
$z=-1$ and for all $\lambda > 1/4$ the shape is nonconvex, 
whilst at $\lambda = 1/2$ we have ergodicity and also nonconvex geometry.
Technically, in all cases at various $\lambda$ 
we have calculated all states by applying the BIM, but then, for
technical reasons compared only the odd states with their exact value, which
are supplied by the conformal mapping diagonalization 
technique\cite{Rob84,PRob934}

In Figs. 8(a-f) we thus observe that there is no clear relationship 
between the
value of $\alpha$ and the degree of classical chaos in all the various
billiards. However, it is an interesting "\,experimental\," result that 
the power law (\ref{eq:power-law}) seems to be universally valid,
with nonuniversal numerical value of $\alpha$ and $A$.

Having established the validity of the power law (\ref{eq:power-law}) it is 
now
most interesting and also immensely CPU time consuming  (it took almost one
month of CPU time on a Convex C3860 to produce Figs. 9 and 10) 
to look at the variation of $\alpha$ with the billiard
shape parameter $\lambda$, which is shown in Fig. 9.
There is a flat
region of almost constant $\alpha$ within $0 \le \lambda \le 1/4$:
It fluctuates slightly around $3.5$.
At $\lambda > 1/4$ the nonconvexities of the 
boundary appear; unlike naive expectation, $\alpha$ now even increases up to 
a value of slightly larger than 5 reached at $\lambda\approx 0.35$ and then
starts to decrease rapidly down to the value of about 2 at $\lambda=1/2$.
Therefore, there is no clear correlation with either the nonconvexities or
the classical chaos.

In addition to $\alpha$ in Eq.(\ref{eq:power-law}) we would also 
like to know the
value of the constant $A$ (the pre-factor) in each case. This is given in 
Fig. 10 by fixing $b=12$ and plotting the mean absolute value of the error 
(averaged over the lowest 100 odd states) versus $\lambda$. Here we see 
that 
the mean error $\langle|\Delta E|\rangle$ is almost constant up to $\lambda 
\le 0.35$ and is equal to about $6 \times 10^{-6}$.
At $\lambda \ge 0.35$ we observe the rapid increasing of 
$\langle|\Delta E|\rangle$. Here again we cannot draw a clear conclusion 
but only note that
the error starts to increase rapidly in the regime where classical chaos 
becomes "\,hard\," (the Kolmogorov entropy increases steeply\cite{Rob83}).

Most of our results are summarized in Table III for three classes of billiard
systems with different type of classical dynamics, namely, integrable, 
KAM-type,
and ergodic systems. We show the calculated values of $\alpha$ and also the
average absolute value of the error $\langle|\Delta E|\rangle$ with fixed 
value of $b=12$.
The table clearly demonstrates that the power law (\ref{eq:power-law})
for the BIM is universal, but not the exponent $\alpha$ and the prefactor 
$A$. It also demonstrates that
classical dynamics has little effect on $\alpha$, whereas the 
nonconvexities of
the boundary might be more important. There is no theory for $\alpha$ so far
except for the circle billiard, for which Boasman\cite{Boas94} predicts 
$\alpha =3$, which might be marginally compatible with our data. 
The bizzare behavior of $\alpha$ in various dynamical regimes reminds us of 
the difficulties in theoretical predictions of, e.g., classical correlation
functions\cite{GGalla94}.
In the table we include also the results of the 
integrable case of the
rectangular equilateral triangle (half of the unit square) where
$\alpha=3.28$ is quite large and the ergodic case of the $1/4$ Heller's 
stadium 
($2\times 2$ square plus two semicircles with a unit radius) in which case
$\alpha =3.0$ is also quite large. 

When thinking about improving the efficiency and the accuracy of the BIM we 
have
also tried a more sophisticated version of the BIM, where we have explicitly 
used a Gaussian integration on the boundary when discretizing our main 
equation (\ref{eq:BIM2}). However, this experience has been negative after 
many careful
checks in various billiards and therefore we decided to resort to the
primitive discretization of the BIM, which is exactly the same approach as 
in \cite{BW84}.
             
\section{Discussion and conclusions}

We believe that our present paper presents quite firm numerical 
(phenomenological) evidence for the relevance of classical chaos for the
effectiveness of the PWDM as a quantal numerical method to solve a quantum 
billiard, which is manifested especially in the semiclassical limit and might
and should be explained in terms of an appropriate semiclassical theory. 
Qualitatively, the reasons for this phenomenon are explained before. The 
parameter $\rho_2$, the fractional volume of
the chaotic component(s), definitely plays an important
role, but is not the only aspect of classical chaos controlling the 
behavior of
the error \dE~ at fixed discretization density $b$. Namely, even in 
rigorously
ergodic systems where $\rho_2=1$ the error \dE~ might be controlled 
by the
slow diffusion in the classical phase space (diffusive ergodic regime, soft
chaos). If the classical diffusion time is much longer than the break time 
~$t_{break}$ ~ ($t_{break} = \hbar/D$, where $D$ is the mean energy level 
spacing) then the
quantal states will be strongly localized in spite of the formal ergodicity
(for a demonstration see Ref.\cite{LiRob95a} for the Robnik billiard and 
Refs.\cite{BCL96,BCHL97} for the stadium billiard) and
therefore they mimic a certain amount of regularity, enabling a better 
accuracy of the
PWDM, i.e., \dE~ is smaller than for completely extended chaotic high-lying
eigenstates where, according to our experience, somehow \dE~ typically 
saturates at
about a few percent of the mean level spacing, even if we drastically 
increase $b$ beyond any reasonable limits. 
Indeed, as can be seen by a comparison of Figs. 2--4, in the case of the 
stadium
this saturation value of \dE is about $10^{-4}$, which is almost two orders of
magnitude smaller than in the Sinai billiard (Fig. 4) 
and the cardioid billiard (Fig. 2).  We think that this is due to the strong
localization of eigenstates in the stadium, which is very well known to
display an unusual abundance of scars\cite{Heller84,Li97,LiHu97}.
Thus our present work is a motivation for a semiclassical theory to explain
this aspect of quantum chaos that exhibits some algorithmic properties of the
PWDM in applying it to quantum billiards with a variety of classical 
dynamics.

Further, we have investigated the 
behavior of the boundary
integral method  with respect to the density of discretization $b$ as
defined in Eq. (\ref{eq:defb}) ($b$ is the number of numerical nodes per de
Broglie wavelength along the boundary) since we expected some relevance of 
nonconvexities and possibly of classical chaos.  
In all cases we discovered that there is a power-law behavior described in
Eq. (\ref{eq:power-law}). We wanted to verify whether there is
any systematic effect of classical dynamics of quantum billiards on $\alpha$
and $A$. The answer is negative. On the other hand, we found that the role of
nonconvex geometry of the boundary might be more important, although no final
conclusion is possible at this point.
The difficulties of the BIM might be expected as 
explained before, and the easiest way to see that is to consider
the semiclassical limiting approximation of the BIM\cite{LiRob95c}.
After explaining two
systematic errors in the literature where the integral the BIM equation is 
derived
and where luckily the two errors mutually compensate for each other 
exactly, we 
have given the correct ({\em regularized}) derivation and discussed
the BIM formalism thus derived. We agree that even in nonconvex 
geometries it
is formally right, but nevertheless practically might be less efficient,
which is expected by 
considering the semiclassical limit mentioned above. 
Therefore, we suggest a generalization
of the BIM by using a multiple reflection (collision) expansion
in calculating the most appropriate Green function, which is
another subject of our current investigation\cite{LiRob95c} and is
important not only for studies in quantum chaos but also in engineering
problems\cite{Baner94} since it would lead to a better 
efficiency of the BIM, namely, larger $\alpha$. 
Most of our results are summarized in Table III, giving the evidence for the
above conclusions. 

It remains an interesting and important theoretical problem to study the
sensitivity of the eigenstates (eigenenergies and wave functions) on the
boundary data of eigenfunctions, of which one aspect is also the dependence of
the eigenstates on the billiard shape parameter. If such sensitivity correlates
with classical chaotic dynamics and at the same time manifests itself in the
accuracy of the purely quantal numerical methods, then such a behavior 
would be
one important manifestation of quantum chaos. This interesting line of thought
in the search for another aspect of quantum chaos has been further 
developed in another work\cite{LiRob96b}, where we also present 
detailed studies of
a level curvature distribution  and other measures of the sensitivity of the
eigenstates.

\section*{Acknowledgments}
We thank Dr. Holger Schanz and Professor Uzy Smilansky for the 
table 
of the eigenenergies for the Sinai billiard, and Dr. Toma\v z Prosen for 
supplying the eigenenergies for the Robnik billiard. We especially thank  
Dr. Thomas Hesse (University of Ulm) for his results and analysis of the
annular billiard, and for many useful discussions.
We also thank Dr. Vladimir Alkalaj, the director of the National Supercomputer
Center, Slovenia, for kind support.
The financial support from the Ministry of Science and 
Technology of the Republic of Slovenia is gratefully acknowledged. This work 
was supported in part by  the grants from Hong Kong Research Grants 
Council (RGC) and the Hong Kong Baptist University Faculty Research 
Grants (FRG).

\newpage
{\bf TABLE I.}  Test of the orthogonality of the eigenstates and the scalar 
product of two consecutive normalized wave functions $O_{12}$, namely, the
ground state and the first excited state, for the Robnik billiard
at different shape parameters. The number in brackets 
represents thepower of 10.\vspace{5mm}\\

\begin{tabular}{|c|c|}\hline
&\\
$\lambda$ & $O_{12}$\\[2.5ex] \hline
0   & 2.0[-16]\\[1ex]
0.1 & -5.0[-15]\\[1ex]
0.2 & 7.8[-10]\\[1ex]
0.3 & 4.8[-6] \\[1ex]
0.4 & 5.5[-4] \\[1ex]
0.5 & 1.5[-3]\\[1ex]
 \hline
\end{tabular}

\bigskip
\bigskip

{\bf TABEL II.}  Stationarity test of \dE~  for the Robnik billiard at
$\lambda=0.27$ for the lowest 400 odd eigenstates. \vspace{5mm}\\

\begin{tabular}{|c|c|}\hline
&\\
Average stretch & \dE\\[2.5ex] \hline
1--100   & 1.54[-7]\\[1ex]
101--200 & 2.21[-7]\\[1ex]
201--300 & 2.77[-7]\\[1ex]
301--400 & 2.03[-7] \\[1ex]\hline
1--400 & 2.13[-7] \\[1ex]
\hline
\end{tabular}

\newpage
{\bf TABLE III.} Power-law exponent $\alpha$  and the average absolute 
value
of the error $\langle|\Delta E|\rangle$ with $b=12$ for different billiards. 
For details of the KAM-type billiards see also Figs. 9 and 10.
\vspace{15mm}\\

\begin{tabular}{|l|l|c|c|}\hline
& & &\\
Type & Quantum billiard  & $\alpha$ & $\langle|\Delta E|\rangle_{b=12}$ 
\\[2.5ex]\hline & & & \\
  & circle (half)  & 2.94 $\pm$ 0.17 & 6.74[-5]\\[1ex]
integrable & circle (full)  & 3.44 $\pm$ 0.18 & 5.97[-6]\\[1ex]
& rectangle-triangle & 3.28 $\pm$ 0.29 & 4.08[-5]\\[1ex]
& segment annulus & 2.23 $\pm$ 0.24 & 1.77[-3]\\[1ex]\hline
& & &  \\
& Robnik (full) & & \\ [-.3ex]
KAM &($0< \lambda < 1/4)$ & $\approx 3.4$ & $\approx$ 5.0[-6]\\[1.2ex]
& Robnik (half) & & \\ [-.3ex]
& ($0< \lambda < 1/4$) & $\approx$ 2.9 & $\approx$ 7.0[-5]\\[1ex]
\hline
& & &\\
&  stadium (1/4) &   3.00 $\pm$ 0.16 & 1.18[-4]\\[1ex]
 & cardioid (full) & 2.10 $\pm$ 0.13 & 3.04[-4]\\[1ex]
ergodic& Sinai (1/4) & 2.47 $\pm$ 0.05  & 3.39[-3]\\[1ex]
& Robnik (full) &  & \\[-0.3ex]
& ($0.3 < \lambda < 1/2$) & see Fig. 9 & see Fig. 10\\[1ex] \hline
\end{tabular}

\newpage
                      
\begin{figure}
\caption{
Geometry of the boundary of the three
desymmetrized billiards: (a) the Robnik billiard, (b) the
Bunimovich stadium, (c) and the Sinai billiard.  
}
\end{figure}

\begin{figure}
\caption{
Ensemble-averaged (over 100 lowest odd eigenstates) absolute 
error (measured in units of the mean level spacing) \dE~ (in logarithmic 
units) versus the billiard shape
parameter $\lambda$ for the Robnik billiard with a fixed density of boundary 
discretization $b=12$. The numerical points are denoted by $\bullet$, which
are joined by a line just to guide the eye.
}
\end{figure}

\begin{figure}
\caption{
Ensemble-averaged (over 100 lowest odd-odd eigenstates) absolute 
error (measured in units of the mean level spacing) (in logarithmic units) 
versus the billiard shape
parameter $a/R$ for the Bunimovich stadium with a fixed density of boundary 
discretization $b=12$. The numerical points are denoted by $\bullet$, which 
are joined by a line just to guide the eye. 
} 
\end{figure}

\begin{figure}
\caption{
Ensemble-averaged (over 100 lowest eigenstates)
absolute error (measured in units of the mean level spacing) 
versus the billiard parameter  $R$ (the radius of inner circle) for 
the desymmetrized Sinai billiard with a fixed density of boundary 
discretization $b=12$. The numerical points are denoted by $\bullet$, which 
are joined by a line just to guide the eye. } 
\end{figure}

\begin{figure}
\caption{
Absolute error of eigenstates (in logarithmic units) versus eigenenergy for
the lowest 400 odd eigenstates of the Robnik billiard at $\lambda=0.27$.
The averages over consecutive stretches (of 100 states each) are given in 
TABLE II, demonstrating that \dE~ is quite stationary. 
} 
\end{figure}

\begin{figure}
\caption{
Notation of the angles and chords used in boundary integral method (BIM).
}
\end{figure}
    
\begin{figure}
\caption{
BIM error (measured in units of the mean level spacing) of eigenstates
versus energy. The error is the difference between 
the BIM value and the exact value. About 95\% of the lowest 1000 
odd states are shown. Plot 
(a) is for the full circle billiard and (b) for the cardioid billiard 
($\lambda =1/2$). In both cases $b$ is fixed, $b=6$.
}
\end{figure}                       

\begin{figure}
\caption{
Ensemble-averaged (over about 95\% of the lowest  100 odd eigenstates) 
absolute BIM error
versus the density of boundary discretization $b$ and the best power-law 
fit for the various billiards. $\bullet$ represents the numerical data and
the curve is the best power law fit 
whose $\alpha$ and coefficient $A$ are given in each box. In (a) we 
have a full circle billiard, in (b) the
$3\pi/2$ segment of a circular annulus with inner and outer radia 0.45 and 
0.5,
respectively, in (c) and (d) the Robnik billiard with $\lambda=0.15 and 1/2$,
respectively, in (e) the 1/4 $2\times 2$ Bunimovich stadium, and in 
(f) the 1/4 Sinai billiard with radius 1/2 inside a square of size 2. 
}
\end{figure}
                       
\begin{figure}
\caption{
Power-law exponent $\alpha$ versus the billiard shape 
parameter $\lambda$ for the Robnik billiard. The error bars denote the 
standard deviation from the best fit.}
\end{figure}

\begin{figure}
\caption{
Ensemble-averaged absolute BIM error (in units of the mean level 
spacing)  $\langle|\Delta E|\rangle$  (in logarithmic units) with fixed 
$b=12$ 
against the biliiard shape parameter $\lambda$ for the Robnik billiard.
}
\end{figure}


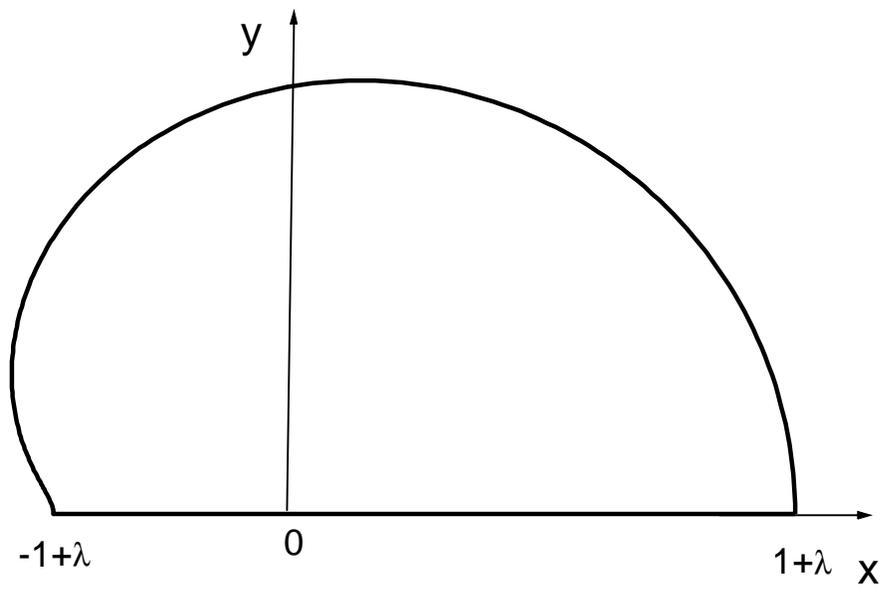
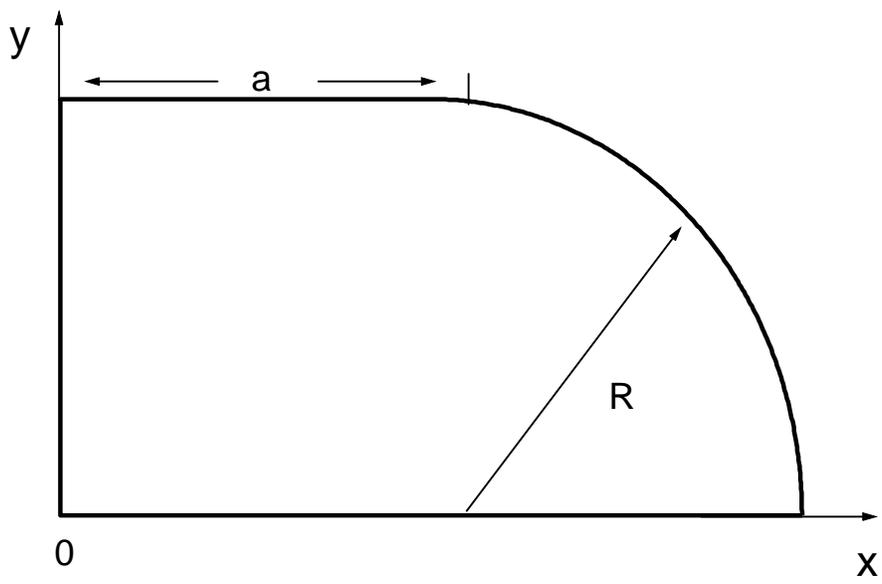
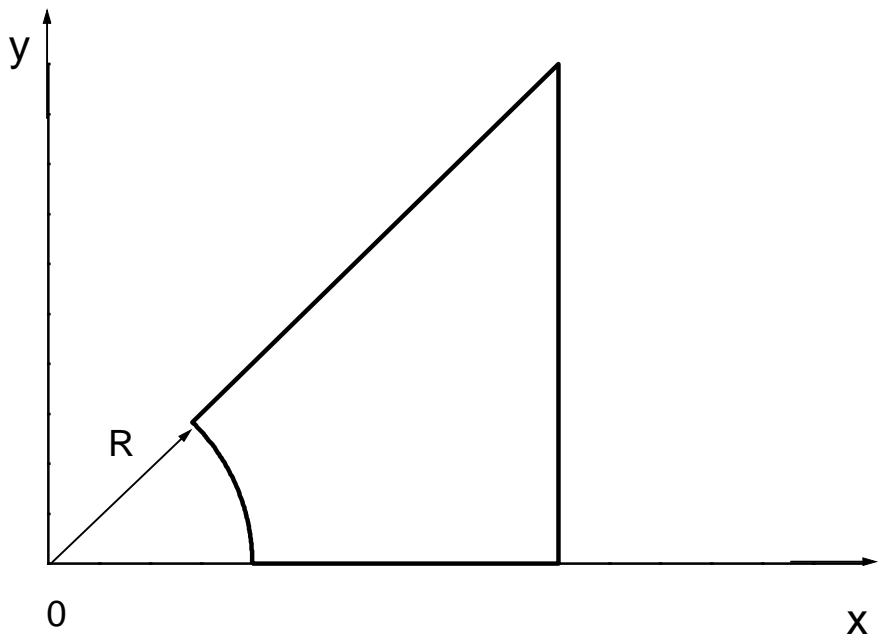
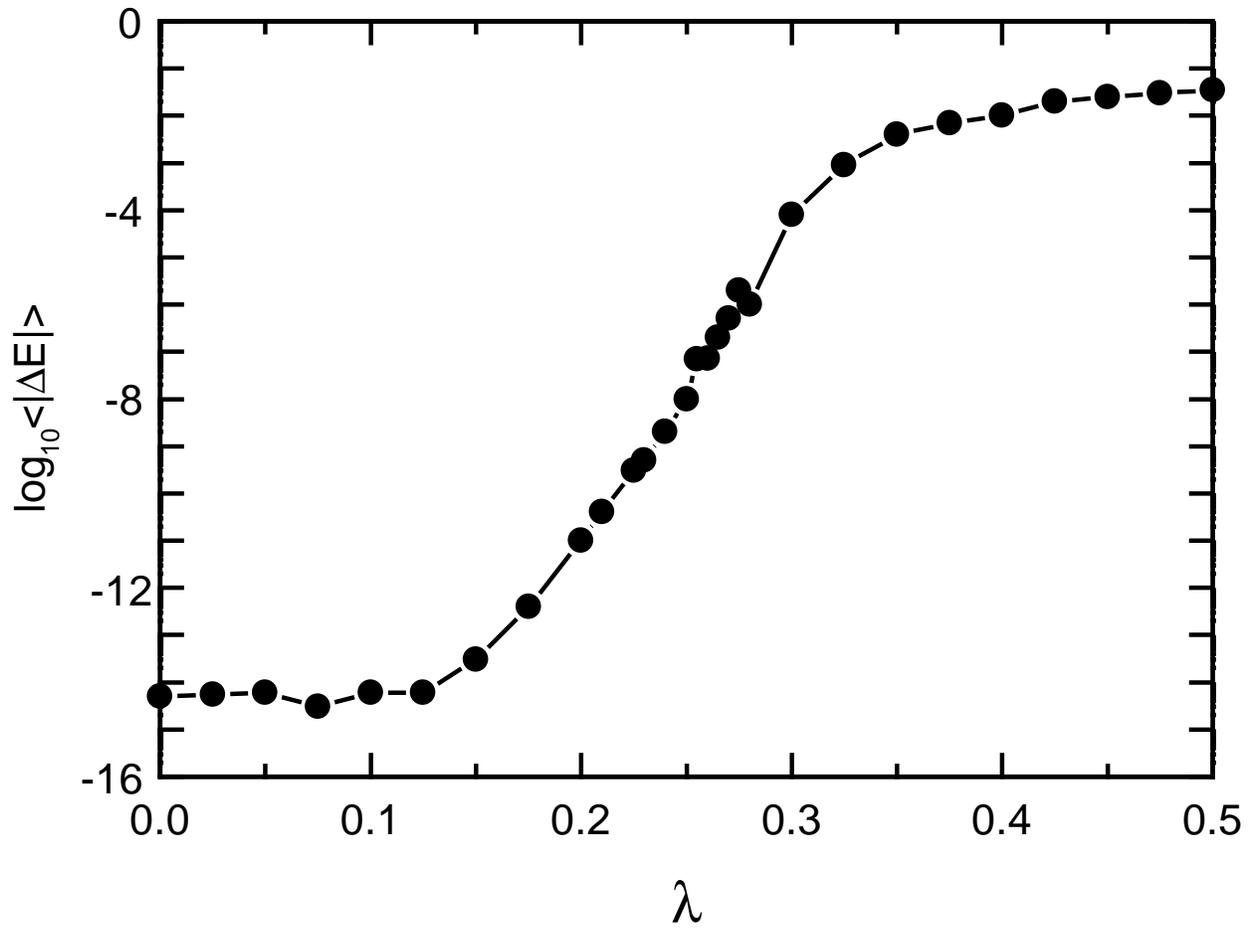
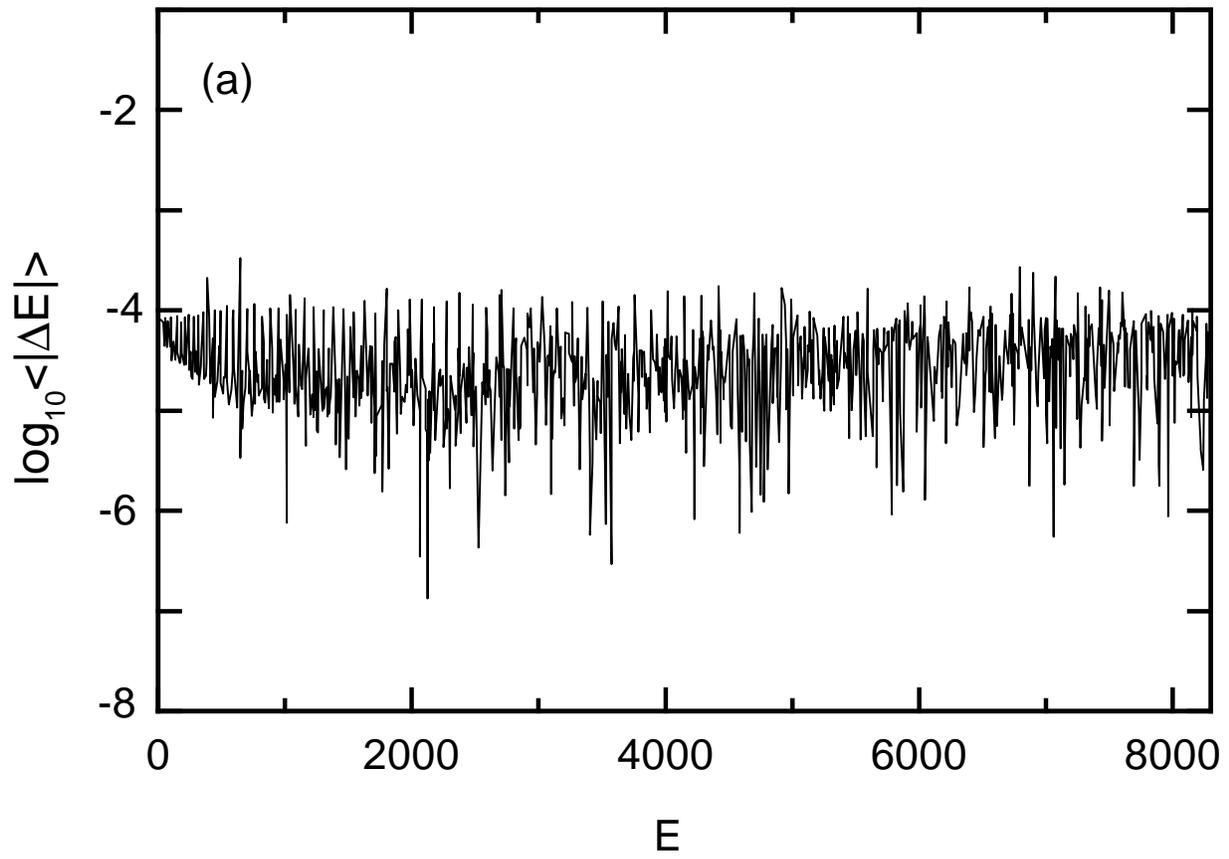
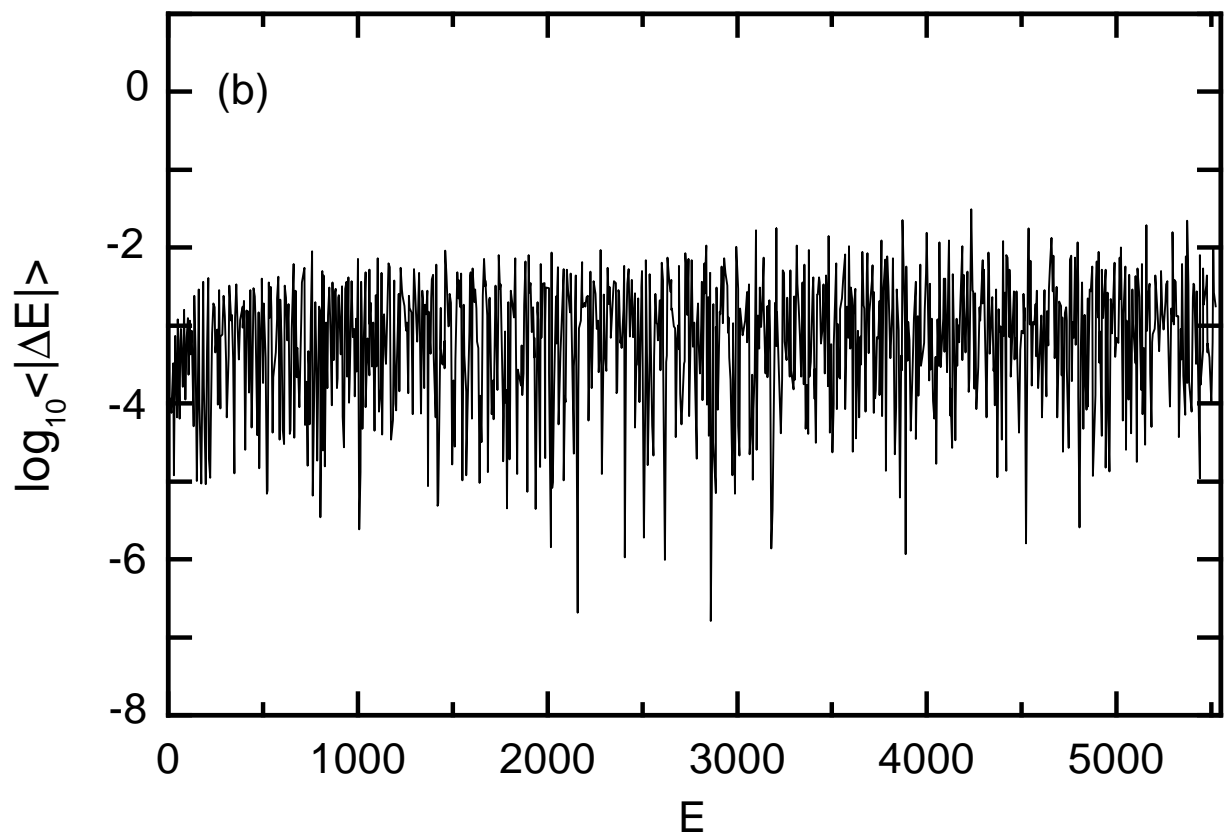
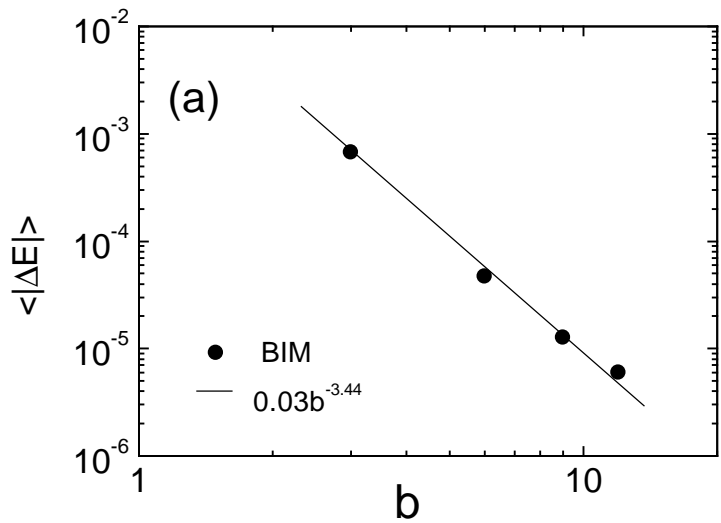
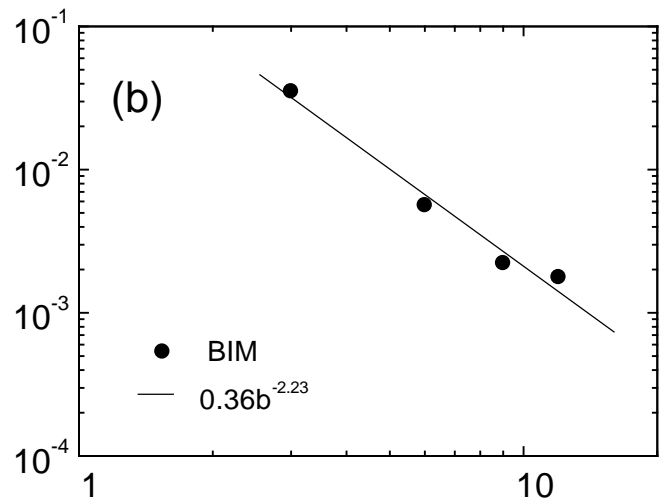
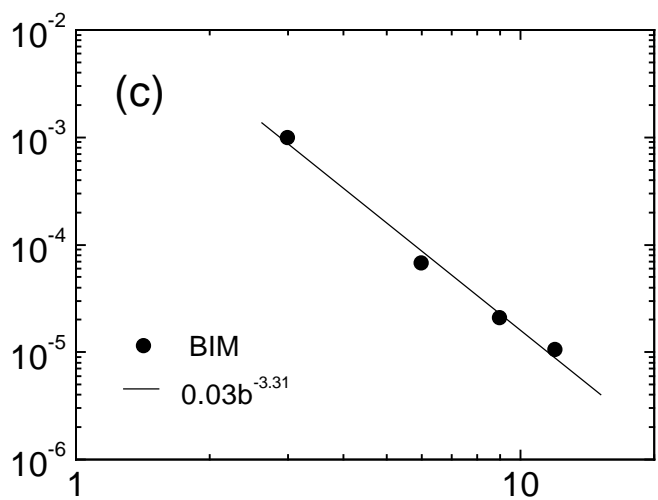
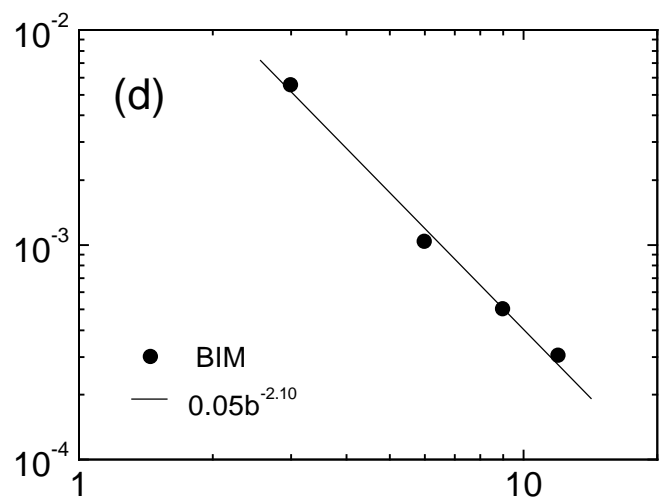
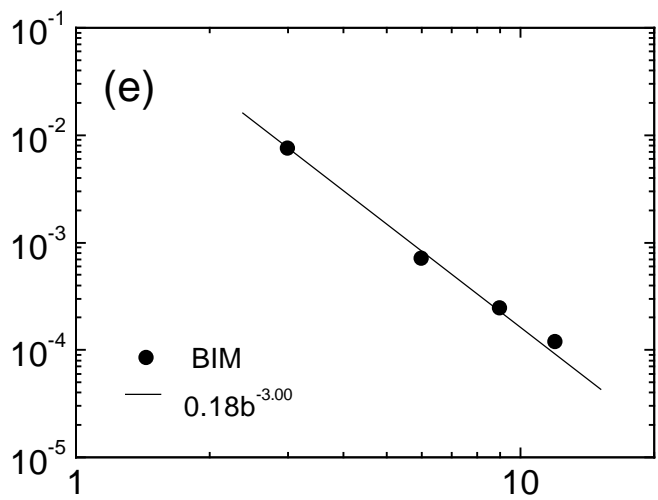
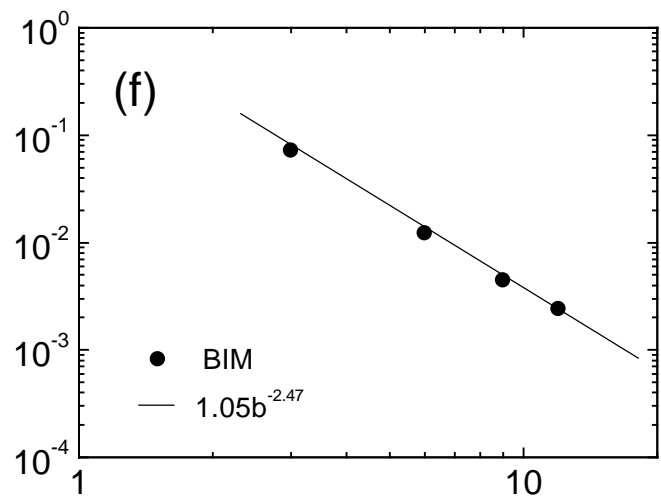
\end{document}